\documentclass[12pt]{article}
\usepackage{latexsym,epsfig}
\usepackage{color}
\usepackage{graphicx}
\usepackage{pifont}
\usepackage{verbatim}
\def\sk{\smallskip}

\def\bg{\bigskip}
\def\beq{\begin{eqnarray}}
\def\eq{\end{eqnarray}}
\def\nl{\noindent}

\begin{document}
\begin{center}
{\bf \large Four Leptons Production in}
\bg 

{\bf \large  $e^-$ $e^+$ Collisions   from 3-3-1 Model }
\vskip .8 cm 

{   E. Ramirez Barreto,  Y. A. Coutinho}

{ Universidade Federal do Rio de Janeiro}
\vskip .5 cm
 
{J. S\'a Borges}

{  Universidade do Estado  do Rio de Janeiro}
\vskip .5 cm

{ Rio de Janeiro - Brazil}
\end{center}

\begin {abstract}
 We study the process $e^- + e^+ \rightarrow e^+ + e^+ + e^- + e^-$ for ILC and CLIC energy regimes, in the minimal 3-3-1 model framework. The main contributions, for the  production of  two  bileptons each one decaying into two same-sign leptons, come   from  s-channel annihilation  ({\it via} $\gamma, Z$ and ${Z^\prime}$) and t-channel electron exchange.  
 We evaluate the number of events for an extra neutral gauge boson mass $M_{{Z^\prime}}$ in the  range  $1$ TeV to $3$ TeV. We compare some distributions from 3-3-1 model with the Standard Model background in order to extract a possible signature of the contribution of bilepton and ${  Z^\prime}$  for the process. From our analysis we conclude that our results  give clear signals  for physics beyond the Standard Model in e$^-$ e$^+$ colliders.   
\end{abstract}


elmer@if.ufrj.br, yara@if.ufrj.br,saborges@uerj.br
\section{Introduction}
The Standard Model (SM) of strong and electroweak interactions is
extremely successful. Its predictions are  consistent with all available experimental data. However, the  new colliders  generation will explore TeV energy regimes  presenting  the possibility of new findings.  On the other hand, the  theoretical extensions of the SM are motivated  by attempting to understand features that are accommodated in the SM but not explained by it, such as the presence of more
than one family in the model and the lack of bounds for the Weinberg angle $\theta_W$.

The 3-3-1 model \cite{PIV,FRA,PHF} predicts the existence of new heavy particles (fermions and bosons), and  it offers an explanation of flavor by anomalies cancellation and clearly shows a limit for the Weinberg angle \cite{ZLI}. The model  offers a possible first step to  understand the flavor question, because the  cancellation of anomalies is obtained by  matching   the number of families and the number of colors. Another feature of the model is that, to keep its  perturbative character, the ratio between the  gauge  couplings forbids 
$\ sin^2 \theta_W  > 1/4$.       

The 3-3-1 model has this name because it is based in the semi-simple gauge group 
${SU (3)_C \times SU (3)_L \times  U (1)_X }$.
It is a gauge theory based in a largest group of symmetry and,  as a consequence,  it predicts the existence of more particles than the SM:  exotic quarks, new  gauge bosons, neutral ($ Z^{\prime}$), bileptons ($V^{\pm }$ , $Y^{\pm \pm}$) and a large number of Higgs. 
In the model, total lepton number is conserved but the separate
flavors lepton numbers are violated.
  This leads to  dramatic signature for double charged  bileptons because they can decay into two same-sign leptons. Such evidence will be accessible by next
generation of $e^-\  e^+$ colliders with center of mass energy: $\sqrt s = 1$ TeV ILC \cite{ILC} and $\sqrt s = 3$ TeV, CLIC \cite{CLI} (CERN).
In this work we will review the minimal version of model in the next section. Our results are presented in the section 3 followed by our conclusions in section 4. 

\section{The Model}
 The  model  has  five additional gauge bosons beyond the
SM ones. In its minimal version they are:  a neutral {$    Z'$} and four heavy charged  bileptons,
 ${  Y^{\pm \pm},V^\pm} $ with lepton number
{$  L = \mp 2$} { (hence their  name)}.
For each generation,  the  usual doublet-singlet pattern per family of SM is replaced by
one triplet of electroweak $SU(3)_L$. 

\begin{eqnarray}
\psi_{L}& = &\left(\begin{array}{c} \nu_{\ell}\\ \ell \\ \ell^{c}
  \end{array}\right)_{L} 
  \sim \left(1, 3, 0\right), 
\end{eqnarray} 

\nl where $\ell^{c}$ is the charge conjugate of $\ell$ ($e$, $\mu$, $\tau$) field and in parenthesis are respectively  the dimensions of the group representation of $ SU (3)_C $,  $  SU(3)_L$ and $ U(1)_X$ charge.

The first quark family, 
\begin{eqnarray}
Q_{L1}& =& \left(\begin{array}{c} u \\ d \\ J_1
\end{array}\right)_{L} \sim \left(3,+2/3, 0\right) 
\end{eqnarray}
\nl transforms as a triplet under $SU(3)_L$ group, and the two other families,
\begin{eqnarray}
Q_{L2}& =& \left(\begin{array}{c} J_2 \\ c \\ s
\end{array}\right)_{L} \sim \left(\bar 3, -1/3, 0\right)  ,\,\
Q_{L3} = \left(\begin{array}{c} J_3 \\ t \\ b
\end{array}\right)_{L} \sim \left(\bar 3, -1/3, 0\right)
\end{eqnarray}
\nl as anti-triplets.
$J_1$, $J_2$ and $J_3$ are exotic quarks with respectively $5/3$, $-4/3$ and $-4/3$ units of positron charge.

The minimum Higgs structure  necessary for symmetry breaking and
that gives quarks and leptons acceptable masses are (three triplets and one anti-sextet):
\beq
\eta =\left(\begin{array}{c} \eta^{0} \\ \eta_{1}^{-} \\ \eta_{2}^{+}
  \end{array}\right)
  ,\,
\rho =\left(\begin{array}{c} \rho^{+} \\ \rho^{0} \\ \rho^{++}
  \end{array}\right) 
  ,\, 
\chi =\left(\begin{array}{c} \chi^{-} \\ \chi^{--} \\ \chi^{0}
  \end{array}\right)
  , \,
S =\left(\begin{array}{llll}
 \sigma^{0}_{1} & \textit{h}^{+}_{2} & \textit{h}^{-}_{1} \\ \textit{h}^{+}_{2} & \textit{H}^{++}_{1} & \sigma^{0}_{2}  \\ \textit{h}^{-}_{1} & \sigma^{0}_{2} & \textit{H}^{--}_{2}  \end{array}\right)
\eq

The breaking of 3-3-1 group to the SM one is accomplished because the neutral component of scalars   acquires  vacuum expectation value (VEV). These scalars  will produce the following hierarchical symmetry
breaking
 $${SU_L(3)\otimes U_X(1)}\stackrel{<\chi>}{\longrightarrow}{    SU_L(2)\otimes
U_Y(1)}\stackrel{<\rho,\eta, S>}{\longrightarrow}{ U_{e.m}(1).}$$
The consistency of the model with SM phenomenology is imposed by fixing a large scale for the VEV of the neutral $\chi$ field ($v_\chi \gg v_\rho, v_\eta$).

One of the main features of the model comes from the relation between the  $SU_L(3)$  
coupling, $g$,  and $U_X(1)$ coupling, $g^\prime$ ($g^\prime/g =t\equiv \tan \theta_W$) that fixes $\sin^2 \theta_W < 1/4$ and is  expressed as:

\begin{equation}
\frac {g^{\prime\, 2}}{g^2} =\frac{\sin^2 \theta_W}{1\, -\, 4 \sin^2 \theta_W}.
\end{equation}

The gauge bosons are $W^{a}_{\mu}$ ($a=1... 8$) in a octet representation of  ${SU(3)_{L}}$ and a singlet $B_{\mu}$ of $U(1)_{X}$.   
The charged and neutral  gauge bosons are defined from the combinations:
\beq
&&W^{\pm}_{\mu}\equiv \frac{W^{1}_{\mu} \mp iW^{2}_{\mu}}{\sqrt{2}},\ 
{    V^{\pm}_{\mu}}\equiv \frac{W^{4}_{\mu} \pm iW^{5}_{\mu}}{\sqrt{2}},\ 
{    Y^{\pm\pm}_{\mu}}\equiv \frac{W^{6}_{\mu} \pm iW^{7}_{\mu}}{\sqrt{2}},
\eq
\beq
A_{\mu}&=& h(t)^{-1/2} \left[ \left(W^{3}_{\mu} - \sqrt{3}\, W^{8}_{\mu}\right)t + B_{\mu} \right],\cr
\cr
Z_{\mu}&\simeq& - h(t)^{-1/2} \left[ f(t)^{1/2} W^{3}_{\mu} +
 f(t)^{-1/2}\left(\sqrt{3}\,t{^2}\, W^{8}_{\mu} - t B_{\mu}\right)
 \right],\cr \cr
{Z^\prime_{\mu}}&\simeq & f(t)^{-1/2}\left[W^{8}_{\mu} + \sqrt{3}\, t\, B_{\mu}\right],
\eq
\nl with\,  $h(t)=1+4 t^2$\,  and\,  $f(t)=1+3 t^2$.
\bigskip

The charged gauge boson masses as a function of VEV's are:

\begin{eqnarray}
M^2_{W}=\frac{1}{4} g^{2}\left({v}_{\eta}^{2} + {v}_{\rho}^{2}\right),
M^2_{V}=\frac{1}{4} g^{2}\left({v}_{\rho}^{2} + {v}_{\chi}^{2}\right),
M^2_{Y}=\frac{1}{4} g^{2}\left({v}_{\eta}^{2} + {v}_{\chi}^{2}\right).
\end{eqnarray}

and the neutral gauge masses 

\begin{eqnarray}
M^2_{\gamma}=~0,~M^2_{Z} \simeq ~\frac{g^{2}}{4}(\frac{g^2+ 4g^{\prime 2}}{g^2 +
3g^{\prime 2}})\left(v^{2}_{\eta} +
v^{2}_{\rho}\right),~M^{2}_{Z^{\prime}} \simeq~ \frac{1}{3}(g^{2}+ 3 g^{\prime 2})v^{2}_{\chi}
\end{eqnarray}

The relation between $Z^\prime$ and $Y$ masses \cite{DION,NG}, in the minimal model is:
\beq
&&{M_{Y}\over M_{{Z^{\prime}}}}={{(3-12\sin^2\theta_W)^{1/2}}\over {2\cos\theta_W}}, 
\eq
\nl which will be used in the present work.

For our purpose, let us consider only the  charged current interactions of leptons with bileptons  given by: 
\beq
&&{\cal L}^{CC}=-\frac{g}{\sqrt2}\left(\ell\, ^T\ C \ \gamma^\mu\gamma^5\, \ell
\ {Y^{++}_\mu}\right),
\eq
\nl where $C$ is the charge conjugation matrix. 
\sk

 The neutral interactions follow from  the Lagrangian
\beq
{\cal L}^{NC} = - \bar \ell \, \gamma^\mu \, \ell\  A_\mu  -  \frac{g}{4}\frac{M_Z}{M_W}[\bar
\ell\, \gamma^\mu\ (v_\ell+a_\ell\gamma^5)\ \ell\, Z_\mu+ \bar
\ell\, \gamma^\mu\ (v^\prime_\ell+a^\prime_\ell\gamma^5)\ \ell\, { Z_\mu^\prime}],
\eq
\nl with
$ v_\ell =   -1/h(t),\ a_\ell = 1, \quad
v^\prime_\ell =  -\sqrt{3/h(t)},\quad 
\  a^\prime_\ell= v^\prime_\ell/3, $

\nl Note that for $t^2=11/6$, $v_\ell$ and $a_\ell$ have the same values as the
SM couplings.
\section{Results}
Let us first consider the production of two real bileptons in \break  $e^- + e^+ \longrightarrow Y^{++} + Y^{--}$. For this calculation and in order to show the relevant distributions, we have used  the package CompHep \cite{HEP}.  The contributions are from $\gamma$, $Z$  and  $   Z^{\prime}$ (s-channel) and $e$ exchange (t-channel).  
In Fig. 1 we show the total cross section for some pairs of ${Z^{\prime}}$ and ${Y^{\pm \pm}}$  masses, as a function of $\sqrt s$.
  A resonance peak is observed for  $M_{Z^{\prime}}= 1$ TeV ($\Gamma_{Z^{\prime}}= 164$ GeV) and $\sqrt s= 1$ TeV. For higher masses and widths, and for  $\sqrt s= 3$ TeV, the peak disappears.
\par

 For an annual integrated  luminosity from  ${\cal L}_{int}=  100$ fb$^{-1}$ to $500$ fb$^{-1}$  the number of events with two final bileptons  is  in the range $ 10^{4}$ to  $ 10^{5}$.
 There is no  similar process in the SM producing two very  massive gauge bosons. However one can estimate the order of magnitude for four leptons production by considering $e^- + e^+ \longrightarrow Z  + Z$  that, for  ${\cal L}_{int}=  100$ fb$^{-1}$, produces around $ 10^{4}$  events  for $\sqrt s = 1 $ TeV and this value reduces to  $ 10^{3}$  for $\sqrt s= 3$ TeV. 
\vskip -1cm
\begin{figure}[!htb] \includegraphics[height=.5\textheight]{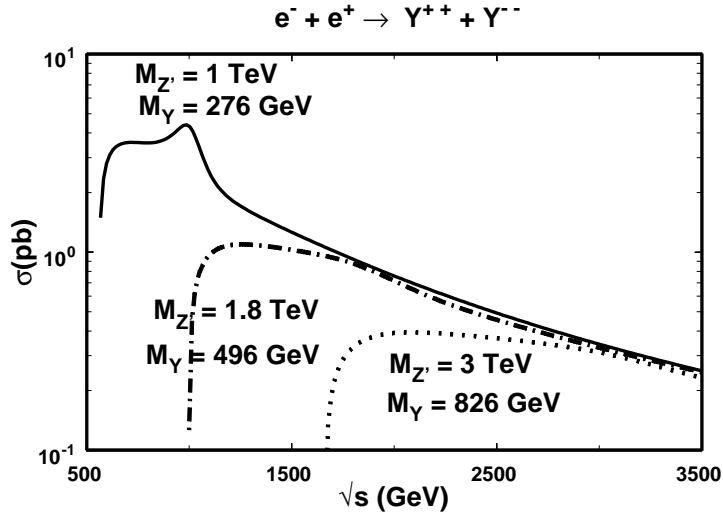}
\vskip -1.2cm
\caption{Total cross section for  $e^- + e^+ \longrightarrow Y^{++} +  Y^{--}$ {\it versus} $\sqrt s$.}
\end{figure}

Another indication of the presence of heavy boson exchange is the forward-backward asymmetry (A$_{FB}$), obtained from  the angular distribution of  $Y^{++}$ relative to the direction of the initial electron, shown in Fig. 2.
 We note again the resonant peak, originated from
${Z^{\prime}}$ exchange, only for $M_{Z^{\prime}} = 1 $ TeV. For higher masses this effect is not present because $Z^\prime$ gets broader, as shown in Table I. 

From the $Y^{++}$ transverse momentum  distribution,  depicted in the Fig. 3,  we observed that the bileptons  are  produced with  $p_t\simeq 50$ GeV for  $M_{Z^{\prime}} = 1 $ TeV and $\sqrt s=1$ TeV. This  value  increases to  $\simeq 150$ GeV when $M_{ Z^{\prime}} = 1 $ TeV and $\sqrt s=3$ TeV. This behavior is inherent to heavy particle production. This important observation allows one to introduce  a minimal cut for the sum of transverse momenta for  each pair of same-sign leptons produced when the  bilepton  are off shell 
($p_{i\, t} + p_{j\, t} \simeq 50$ GeV) for $\sqrt s=1$ TeV and $150$ GeV for $\sqrt s=3$ TeV.  

\vskip -1cm
\begin{figure}[!htb] 
 \includegraphics[height=.5\textheight]{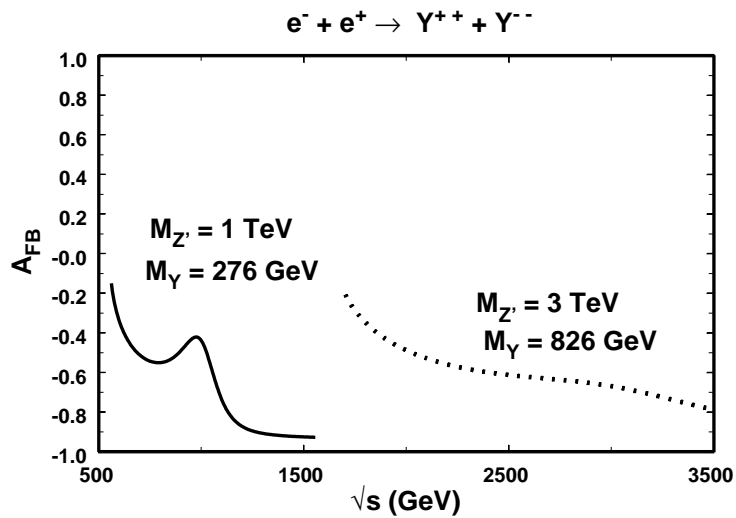}
\vskip -1.2cm
\caption{The forward-backward asymmetry extracted from the angular distribution of one bilepton  relative to the direction of the initial electron {\it versus} $\sqrt s$.}
\end{figure}
\newpage

This process was also calculated in \cite{LONG}, where the authors only calculated total cross section arriving to same order of magnitude than we obtained. In addition, in the present paper, we calculated distributions: that allow one to introduce kinematic cuts to be used in bilepton off-shell calculation and show a signature for heavy neutral boson exchange.
 
Next we consider the complete process $e^- + e^+ \rightarrow e^+ + e^+ + e^- + e^-$, experimentally accessible, not yet calculated in literature, for which the gauge bosons are off-shell. In this case the number of  tree diagrams  increases from 36 corresponding to SM contributions to  133   diagrams  for 3-3-1 model. Notice that  97 diagrams involve  $ Z^{\prime}$,  $ Y^{\pm \pm}$ and their combinations. 
Due to the calculation complexity  to obtain  cross sections and distributions it is mandatory to  use again CompHep package\cite{HEP}.
\par

We adopted,  for the detector  acceptance, an angular cut of $\vert \cos \theta \vert \le 0.995 $ for the direction of final leptons relative to the beam, and energy cut of $5$ GeV for  final leptons \cite{MIK,AZU,L3C}. As we are interested to show a signature for the bilepton existence, we have also selected an invariant mass cut of two same-sign leptons $\vert M_{ee}-M_{Y^{\pm \pm}}\vert \simeq \Gamma_{Y^{\pm \pm}}$, where the $Y^{\pm \pm}$ widths are presented in Table I. 
\vskip -1cm
\begin{figure}[!htb]\includegraphics[height=.5\textheight]{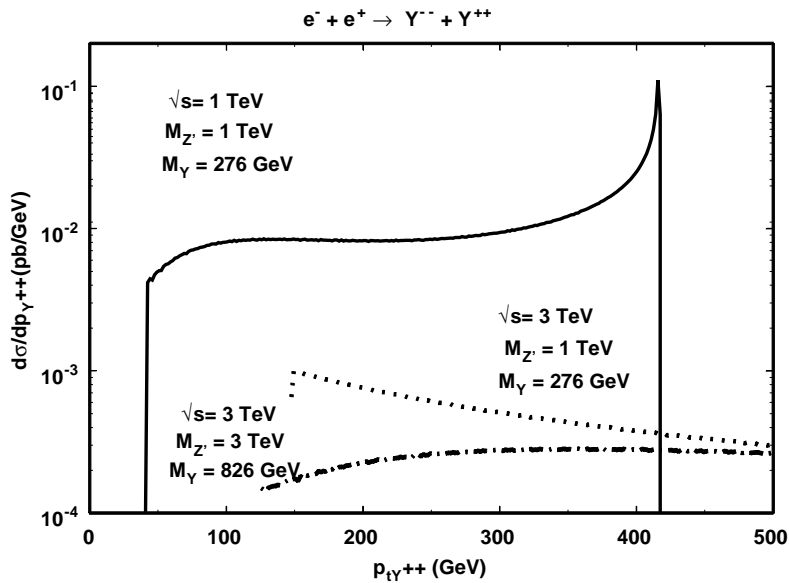}
\vskip -1.2cm
\caption{Final bilepton transverse momentum distribution.} 
\end{figure}
\par
As mentioned before, one is allowed to apply a cut for the sum of transverse momenta of same-sign leptons.
The study of transverse momentum of a final lepton can be done from  Fig. 4 for two values of $\sqrt s$. By this figure we observe that they are mainly produced with high $p_t$, differently from those  produced considering the SM. This is an indication that they are originated from heavy particle decay. 
\par
\vskip -1cm
\begin{figure}[!htb]  \includegraphics[height=.5\textheight]{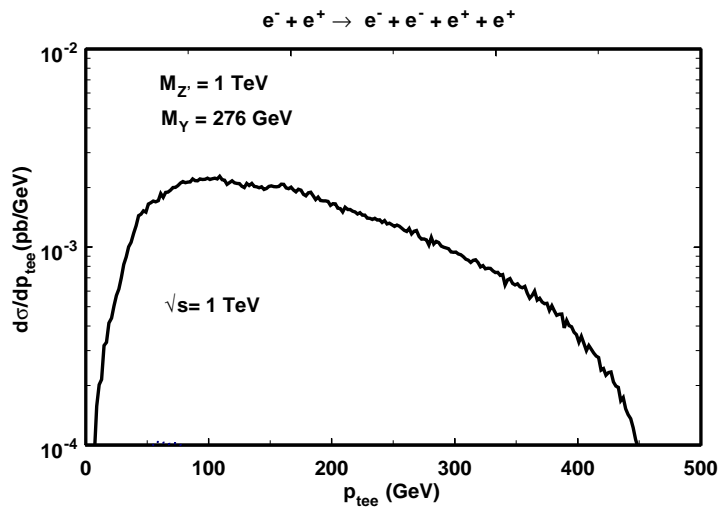}
\vskip -1.2cm
\caption{Final lepton transverse distribution in 3-3-1 model for $M_Y = 276$ GeV and $\sqrt s=1$ TeV.}
\end{figure}

Another interesting distribution that reveals the existence of bileptons is Fig. 5 that shows  the invariant mass distribution of two same-sign leptons ($M_{e e }$) for $\sqrt s=1$ TeV. It is clear that one can identify the bilepton as a resonance by reconstructing it from same-sign final leptons momenta. The same behavior was observed for $\sqrt s=3$ TeV. On the other hand this is not observed in SM that leads to a flat distribution. 
\begin{figure}[!htb]  \includegraphics[height=.5\textheight]{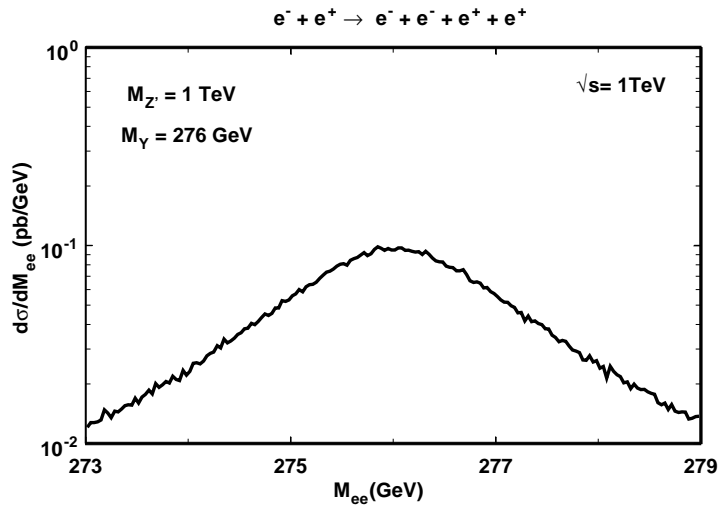}
\vskip -1.2cm
\caption{ Invariant mass distribution of two same-sign leptons in 3-3-1 model for $M_Y = 276$ GeV and $\sqrt s=1$ TeV, a similar distribution was obtained for  $\sqrt s=3$ TeV.}
\end{figure}

The angular distribution between same-sign leptons obtained in 3-3-1 calculation is shown in Fig. 6. For an energy $\sqrt s=1$ TeV, the angle between the same sign leptons is $\simeq 66^{\circ}$,
decreases to $0^{\circ}$ for $3$ TeV, this indicates a slow heavy particle decay.

\begin{figure}[!htb]  \includegraphics[height=.5\textheight]{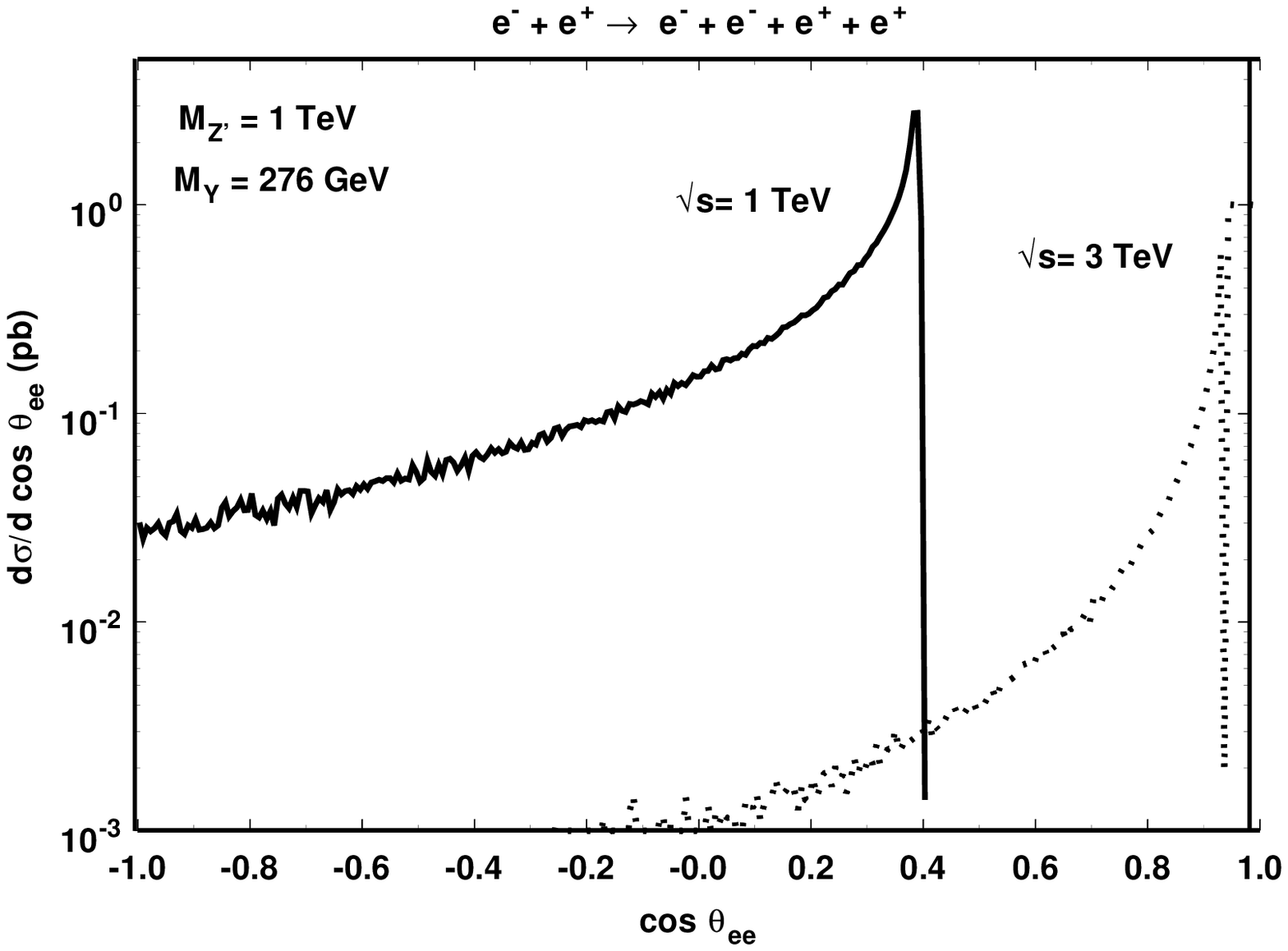}
\vskip -1.2cm
\caption{Same sign leptons angular distribution in 3-3-1 model for $M_Y = 276$ GeV and $\sqrt s=1$ TeV and $\sqrt s=3$ TeV.}
\end{figure}

Finally, in Fig. 7 we present the total cross section as a function of bilepton mass. This picture allows to know the accessible range for $M_Y$, that can be explored in next linear colliders. Considering a luminosity ${\cal L} = 100$ fb$^{-1}/y$, and a range $276$ GeV $ \le M_Y \le 500 $ GeV, we obtained around $10^2$ to $10^4$ events for $\sqrt s =1$ TeV. Increasing $\sqrt s$ to $3$ TeV and exploring up to $M_Y= 800$ GeV, this range turns into $10^3$ to $10^2$ events, what is also a measurable yield.     

3-3-1 model results ($M_{Z^{\prime}}=1$ TeV and  $M_{Y^{\pm\pm}} = 276$ GeV) for total cross section {\it versus} $\sqrt{s}$ are approximately five orders of magnitude bigger than SM.
The unitarity property of the total cross section is assured by  the presence of the extra gauge boson $Z^{\prime}$. 

\begin{figure}[!htb]\includegraphics[height=.5\textheight]{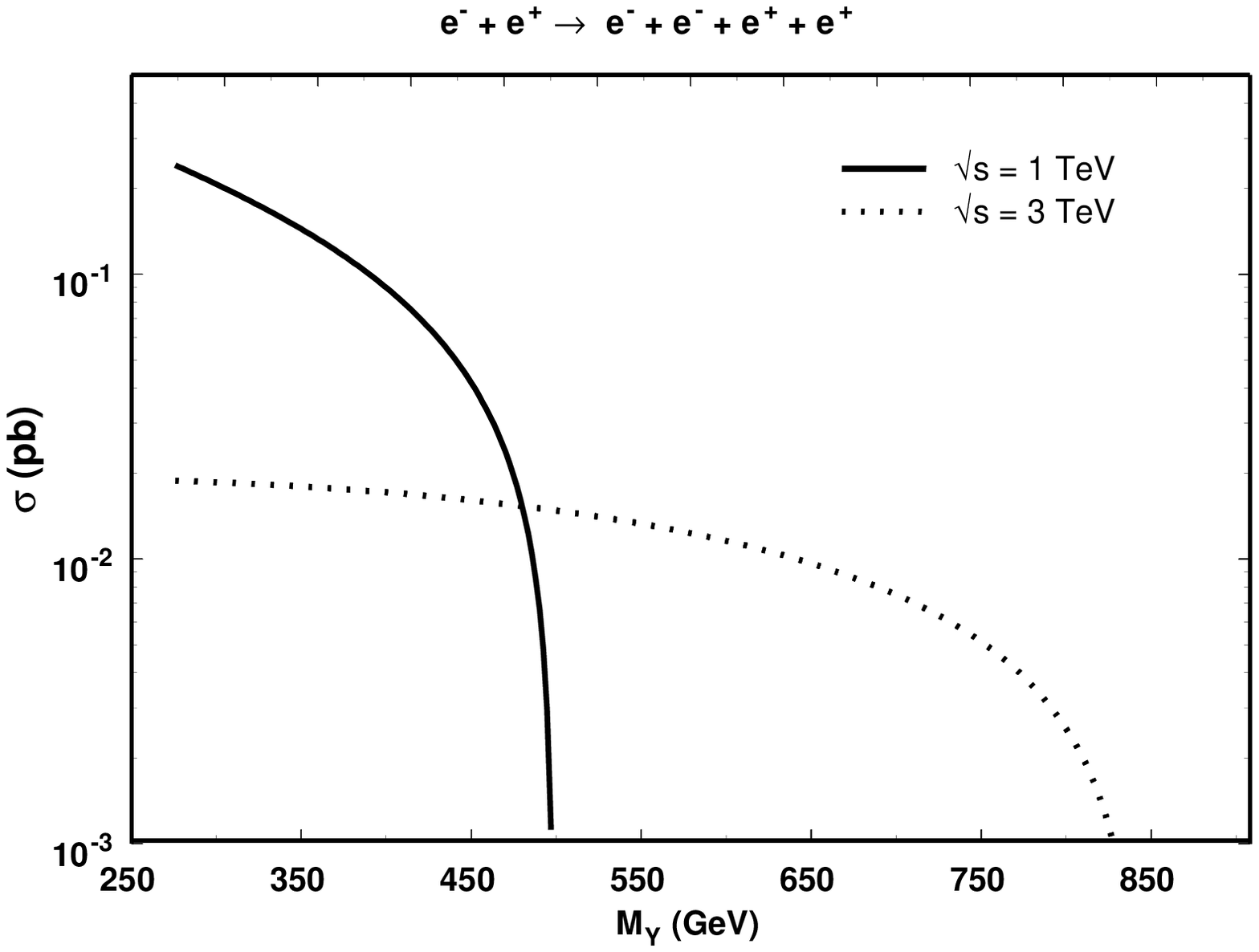}
\vskip -1.2cm
\caption{Total cross section versus $M_Y$ for $\sqrt s= 1$ and $3$ TeV.}
\end{figure}

\section{Conclusion}
We explored the process $e^-  e^+$ at energies corresponding to next generation of linear colliders in order to find signals of the existence of new gauge bosons, predicted by the chiral extension of the SM, called 3-3-1 model.
\par
We have assumed a special constraint between bilepton and neutral gauge boson masses, respecting the experimental bounds that, even being a consequence of the model, it is not often used in the literature.
\par     
In the present paper, to find a signal of new physics, we concentrated our study in four  charged leptons production ($e^- + e^+ + e^-  + e^+$). This process was selected because double charged bilepton mainly decays into two same-sign leptons leading to a non standard  signature.
\par 
We started with a calculation of on-shell bilepton production as a guide to the complete calculation. In literature exist a similar calculation, but only for total cross section.
Instead, our results lead us to implement a cut on the transverse momenta of final leptons when bileptons are off-shell. Another indication that followed from this calculation was a resonance peak in the forward-backward asymmetry related to $Z^{\prime}$ exchange.
\par     
To perform the complete calculation of four leptons production, we used CompHep package. We have applied angular and energy cuts for the detector acceptance. In order to distinguish the signal from SM background, we introduced cuts in invariant mass and transverse momenta of final leptons. We evaluate that the number of events produced by year, for a bilepton mass range from  $M_{Y^{\pm\pm}} = 276$ GeV to $M_{Y^{\pm\pm}} \simeq 830$ GeV, and for $\sqrt{s}=1$ TeV and $\sqrt{s}=3$ TeV, varies from $10^4$ to $10^2$, a promising signature.  
\par 
The next generation of linear colliders working with energy in scale of $1$ TeV and beyond, is a good place to explore a new physics predicted by 3-3-1 model.
\vskip 1cm
                                                                     
We acknowledge the financial support from CAPES (E.R.B.) and FAPERJ (Y.A.C.).

\begin{table}[h]\label{manga}
\begin{center}
\begin{tabular}{||c|c|c|c||}
\hline \hline
&      &      & \\
$M_{Z^{\prime}}$ (GeV) &  $\Gamma_{Z^{\prime}}$ (GeV) & $M_{Y^{\pm\pm}}$ (GeV) & $\Gamma_{Y^{\pm\pm}}$ (GeV) \\
&      &     &   \\ \hline
\hline
&      &      &  \\
$1000$ & $164$ &  $276$  & $2.33$ \\
&      &   &  \\ \hline
\hline
&      &    &    \\
$1800$  &  $788$   & $496$ & $4.2$ \\
&      &    &    \\ \hline
\hline
&      &    &    \\
$2200$ & $998$  & $600$ & $6.4$ \\
&      &    &     \\ \hline
\hline
&      &    &    \\
$3000$ & $1405$  & $826$ & $15.2$ \\
&      &    &     \\ \hline
\hline
\end{tabular}
\end{center}
\caption{Some widths for new gauge bosons $Z^{\prime}$ and bilepton $Y^{\pm\pm}$ in the minimal 3-3-1 model.}
\end{table}

\end{document}